\documentclass[twocolumn,showpacs,preprintnumbers,amsmath,amssymb]{revtex4}
\usepackage[dvips]{epsfig}

\begin{document}

\title{Universality of the Crossing Probability for the Potts 
Model for 
$q=1,2,3,4$}

\author{Oleg A.Vasilyev}

\email{vasilyev@itp.ac.ru}
\affiliation{%
L.D.Landau Institute for Theoretical Physics RAS,
117940 Moscow, Russia }

\date{\today}
\bigskip

\begin{abstract}
The universality of the crossing probability $\pi_{hs}$
of a system to percolate only in the horizontal direction
was investigated numerically by 
a cluster Monte-Carlo algorithm for the $q$-state Potts model for 
$q=2,3,4$
and for  percolation $q=1$.
We check the percolation through Fortuin-Kasteleyn clusters near the 
critical point on the square lattice by using representation of the Potts 
model as the correlated site-bond percolation model.
It was shown that probability of  a system 
 to percolate only in the horizontal direction $\pi_{hs}$ has universal 
form $\pi_{hs}=A(q) Q\left( z \right)$
for $q=1,2,3,4$ as a function of the scaling variable
$z= \left[ b(q)L^{\frac{1}{\nu(q)}}(p-p_{c}(q,L))
\right]^{\zeta(q)}$. Here, $p=1-\exp(-\beta)$ 
is the  probability of a bond to be closed, $A(q)$ is the nonuniversal 
crossing amplitude,
$b(q)$ is the nonuniversal metric factor, 
$\nu(q)$ is the correlation length index,
$\zeta(q)$ is the additional
scaling index.
 The universal function $Q(x) \simeq 
\exp(-|z|)$. Nonuniversal scaling factors were found numerically. 
\end{abstract}


\maketitle

\section{Introduction}

The concept of the universality and scaling relations~\cite{St} is general 
concept
of the modern phase transition theory.
The main point  of the scaling theory is
that in the vicinity of the critical point for a system of linear size 
$L$, the critical behavior
 of thermodynamical quantities
can be expressed as a universal function of 
two variables:  reduced temperature $\tau=\frac{T-T_{c}}{T_{c}}$ and 
external field $h$.
 The finite-size scaling 
of thermodynamical functions of spin models was studied
theoretically and numerically~\cite{FB1,FB2,Landau1,Landau2}.
In Ref.~\cite{PF}, Privman and Fisher proposed the idea of the universal 
finite-size
scaling with nonuniversal metric factors.
For example, for the free-energy density of system size $L$ and dimension 
$d$
\begin{equation}
\label{eq:f}
f(T,h;L)=L^{-d}Y \left(C_{1} \tau L^{\frac{1}{\nu}},C_{2} 
h L ^{(\beta+\gamma)/\nu} \right)
\end{equation} 
where  $\nu$ is 
the scaling index for correlation length, 
$\beta$ is the scaling index of magnetization and
$\gamma$ is the scaling index of magnetic susceptibility.
For  systems of different boundary conditions, aspect ratios 
and geometries (square, honeycomb, triangle), the scaling function
$Y(x,y)$ is universal and only metric factors $C_{1},\;C_{2}$
are system dependent. 
Scaling properties of thermodynamical functions of the Potts model 
were investigated in Refs.~\cite{KL,dQ}.

Langlands, Pichet, Pouliot and Saint-Aubin~\cite{Lang1}
show that for site and bond percolation
on square, honeycomb and triangle lattices with the aspect ratios
$a$, $a\sqrt{3}$ and $a \sqrt{3}/2$ respectively, the
crossing probability $\pi_{h}$ of a system to percolate in the horizontal 
direction is the universal function of $a$.
 Hu, Lin 
show that  by choosing a very small number of
nonuniversal metric factors, all scaled data for percolation functions  
and the number of percolating clusters on  square, honeycomb and triangle 
lattices
may fall on the same universal scaling functions~\cite{Hu1,Hu2}.
Their  scaling argument was $x=(p-p_{c})L^{\frac{1}{\nu}}$ where $p_{c}$ 
is the critical point, $L$ is the lattice size, $\nu$ is the correlation 
length index. 
The scaling of crossing probabilities for the three-dimensional 
percolation
was investigated in Ref.~\cite{Hu3}.

The continuum limit of the crossing probability $\pi_{h}(p_{c})$
was investigated by J. Cardy by conformal field 
methods~\cite{Cardy0,Cardy1,Cardy2}.
The analogous formula for the crossing probability 
$\pi_{hv}=\pi_{h}-\pi_{hs}$
was  found by Watts~\cite{Watts}.
The works of Smirnov~\cite{Sm1,Sm2} analytically proved that the crossing 
formula
holds for the
continuum limit of site percolation on the triangle 
lattice~\cite{Sm1,Sm2}.

The $q$-state Potts model can be represented as 
the correlated site-bond percolation in terms of Fortuin-Kasteleyn (FK)
clusters~\cite{FK}. At the critical point of the second order phase 
transition Potts model,   FK-clusters exhibit the percolation
transition. So there is an intrinsic relationship
between critical properties of the Potts model and percolation properties
of FK clusters.
The universality 
of the crossing probability for 
the Ising model on rectangle lattices of square,
honeycomb and triangle geometries was investigated by 
Langlands {\it et all.}~\cite{Lang2,Lang3}.
 Hu, Chen and Lin  show the universality 
of the crossing probability $\pi_{h}$ and a number of percolation clusters
for the correlated site-bond percolation $q=2$~\cite{Hu4}.

 In this paper the the crossing probabilities
of FK clusters is studied numerically. 
We investigate the universality of the crossing probability
with respect to a number of spin states $q$ of the Potts model.
We show numerically that  
the probability of a system to percolate only in horizontal direction
 $\pi_{hs}$
is an universal function of the scaling variable 
$z=\left[b(q) 
L^{\frac{1}{\nu}}(p-p_{c})\right]^{\zeta(q)}$ 
for 
$q=1,2,3,4$ where $p=1-\exp(-\beta)$ is the probability of bond to be 
closed,
$\beta=1/T k_{B}$ is an inverse temperature, 
$p_{c}=1-\exp(-\beta_{c})$ 
is the  location of critical point in  the $p$ scale, $b(q)$ is a 
nonuniversal metric factor and $\zeta(q)$ is a 
 scaling index, described dependence of the form of the crossing 
probability on $q$.

We show that for each value $q=1,2,3,4$
on the square, lattice the index $\zeta$ is practically 
does not depend on the lattice size.
Therefore, by introducing this index $\zeta(q)$,
we can  lay on the same curve the points  
for different $q$ in the critical region.

\section{ Crossing probabilities  for the  Potts model}

In the Potts model, each spin $\sigma_{i}$ can take one of 
the $q$ different
values $1,\dots,q$
and the  Hamiltonian is written as
${\cal H} = - J \sum \limits_{(i,j)} \delta (\sigma_{i},\sigma_{j} ) $, 
where $J$ is the ferromagnetic coupling constant, which we set it equal 1.
The partition function of the Potts 
model~\cite{Baxter} is
\begin{equation}
\label{eq:potts}
Z= \sum \limits_{\sigma}\exp(-\beta {\cal H} (\omega))=
\sum_{\sigma} \prod \limits_{(i,j)}\left[ (1-p)+p
\delta(\sigma_{i},\sigma_{j}) \right]
\end{equation}
where $\beta =1/T k_{B}$ is the inverse temperature, $p=1-\exp(-\beta)$ 
is the probability of bond to be closed and $1-p=\exp(-\beta)$ 
is the probability of bond to be open;
summation is performed over all spin configurations $\sigma$. 
Sometimes the term in square brackets is 
expressed via $v=\exp(\beta)-1$: 
$ \left[  (1-p)+p \delta(\sigma_{i},\sigma_{j})  \right]
=\exp(-\beta)\left[  1+v \delta(\sigma_{i},\sigma_{j})  \right]$,
but we write Eq~(\ref{eq:potts}) via $p$
to emphasize the fact that in correlated site-bond percolation~\cite{FK},
the probability of bond to be closed is $p$ and to be open is $1-p$. 

 For the square lattice of  linear size $L$ with periodic boundary 
conditions
the total number of bonds is $N=2L^{2}$. Let us define by ${\cal L}$ the 
graph of all the edges on the lattice.
The product over all bonds $(i,j)$ consists of $N$ terms, so the product 
may be expanded as sum of $2^{N}$ terms. 
 Let us associate to each of these $2^{N}$ terms subgraph $G$ of graph 
${\cal 
L}$, by following rule. Each of $2^{N}$ terms can be considered as 
the product
of $N$ factors. Each factor $[(1-p)+ p \delta(\sigma_{i},\sigma_{j})]$ 
corresponds to some edge $(i,j)$ of the graph ${\cal L}$. To construct 
$G$, we perform the following procedure.
If this factor for edge $(i,j)$ is equal to $1-p$, we delete edge $(i,j)$ 
from subgraph.
If this factor for edge $(i,j)$ is equal to $p$,
we leave this edge in the subgraph. So we associate to
each term in sum (\ref{eq:potts}) subgraph $G$. Each subgraph $G$
consists of $b(G)$ edges (closed bonds) and $C(G)$ connected components. 
The term in~(\ref{eq:potts}) corresponding to $G$
contains factor
$(1-p)^{N-b(G)} p^{b(G)}$, and delta-functions guarantee equivalence of
spins in each connected component. As a result of the summation over 
all possible spin configurations, the contribution of
this subgraph $G$ into the partition function is 
$q^{C(G)}(1-p)^{N-b(G)}p^{b(G)}$.
Therefore, we can replace the summation over all spin configurations
in~Eq.~(\ref{eq:potts}) by summation over all possible subgraph $G$
on ${\cal L}$
\begin{equation}
\label{eq:pottsg}
Z=\sum \limits_{G \in {\cal L}} q^{C(G)} (1-p)^{N-b(G)}p^{b(G)} 
\end{equation}
We shall keep in mind that
 this definition is valid even for the non-integer value 
of  $q$. The partition function of the Potts model for  $q \to 1$
corresponds to the percolation, where $p$ is a probability 
of bond to be occupied. 

Now we introduce the crossing probability $\pi_{hs}$, the probability of a 
system to percolate  only in the horizontal direction while
the percolation in the vertical direction is absent.
We must distinguish it from the  probability of a system to 
percolate in the horizontal direction irrespective to percolation in 
the vertical 
direction  $\pi_{h}$. We define the indicator function $I_{hs}(G)$ for
each subgraph $G$ in accordance with the rule
\begin{equation}
\label{eq:ihs}
I_{hs}(G)=\left\{  
\begin{array}{ll}
1 & {\rm if} \;\; G \;\;{\rm percolates \; only \; in \; horizontal 
\; direction}\\
0 & {\rm in \; all \; others \; cases} \\
\end{array}
\right.
\end{equation}
We mean that $G$ percolates only in the horizontal direction
if it contains at least one connected component
 touching left and right sides of the lattice,
and there are no components  joining top and bottom 
sides. Therefore, the crossing probability $\pi_{hs}(\beta;L,q)$ 
of a system of  size $L$ with $q$
possible spin states to 
percolate only in the horizontal direction at $\beta$ can be written as
\begin{equation}
\label{qg:phs}
\pi_{hs}(\beta;L,q)=\frac{1}{Z} \sum \limits_{G \in {\cal L}}
I_{hs}(G) q^{C(G)} (1-p)^{N-b(G)} p^{b(G)}
\end{equation}
Here, $p=1-\exp(-\beta)$.
We can introduce the crossing probability for vertical direction 
$\pi_{vs}$ by the same way. But on the  square lattice 
$\pi_{hs}=\pi_{vs}$ and later on, investigate only $\pi_{hs}$.

Let us notice that $\pi_{hs}(\beta;L,q)$ 
has maximum near the critical point $\beta_{c}(L,q)$ because
in the ordering phase $\beta > \beta_{c}(L,q)$, the most
probable  configurations contain large clusters
touching top and bottom sides of the lattice,
 and do not contribute into $\pi_{hs}$.

\section{Numerical results}
\label{secdet}

We use the Wolff cluster algorithm~\cite{Wolff} to generate the different
spin configurations. For each spin configuration
we generate a bond configuration in accordance with~\cite{FK}. Then, we 
break the lattice into
independent clusters of connected sites by using 
the Hoshen-Kopelman~\cite{HK}  
algorithm. Then,  we analyze crossing  properties
$I_{hs}(G)$ of this configuration.
Between checking  the crossing  we skip 5 Monte-Carlo steps.
For each value of $\beta$, the averaging is performed over 
$10$  series each of  $10^5$ configurations.
 The total number of configurations is $10^6$.
Sets of configurations are
 used for the estimation of the numerical inaccuracy.
Quantity $I_{hs}$ is an indicator function.
It means that it takes values 0 or 1 for each configuration.
Therefore, the resolution of our computations for $\pi_{hs}$ is $10^{-6}$.
We compute data for $q=1({\rm percolation}),2,3,4$
and lattice sizes $L=16,32,48,64,80,96,112,128$.
For each pair  $(q,L)$, we perform computation for 200
values of $\beta$ (or $p$ for percolation) in the critical region.

There is a very important question, how the scaling variable should be 
chosen?

For the percolation $q=1$ the scaling variable is  defined as the 
deviation of
the bond concentration from critical point $r=p-p_{c}$.
For the Potts model, we can take 
\begin{enumerate}
\item $t=\frac{T-T_{c}}{T_{c}}$
\item $\tau=\frac{\beta-\beta_{c}}{\beta_{c}}$
\item $r=(1-\exp(-\beta))-(1-\exp(- \beta_{c})=p-p_{c}$ 
\end{enumerate}
The first variant $t=\frac{T-T_{c}}{T_{c}}$ is usual for the investigation 
of thermodynamic quantities~\cite{St,Baxter}, the second one 
is widely used for the approximation of magnetic susceptibility $\chi$
of the Ising model in the critical region~\cite{Wu},
and the third variant $r=p-p_{c}$ may be argued by the fact 
that for the site-bond correlated percolation, 
the probability of bond to be occupied is $p=1-\exp(-\beta )$ -- see 
equations (\ref{eq:potts}), (\ref{eq:pottsg}).

Let us evaluate the symmetry of the crossing probability as a function
of all these variables
by comparing  third moments of the function $\pi_{hs}$.
We assume that  $\pi_{hs}$ must be the most symmetric function
as a function of "right" variable.
We calculate  four moments  of $\pi_{hs}(x)$ as a functions of 
$x=t,\tau,r$ by numerical integration in accordance with formula
 $\mu_{0} =\int \pi_{hs}(x) dx$, $\mu_{1} =\frac{1}{\mu_{0}}\int x 
\pi_{hs}(x) dx$, 
 $\mu_{k}=\frac{1}{\mu_{0}} \int (x-\mu_{1})^k 
\pi_{hs}(x) dx$.
Results of computations $\mu_{3}$ for the lattice size $L=128$ are given in 
Table~\ref{table:mom}. 
\begin{table}[h]
\caption{The third moment $\mu_{3}$ of the crossing probability 
$\pi_{hs}$,
computed for different variables $t$, $\tau$ and $r$}
\begin{center}
\begin{tabular}{|c|c|c|c|}
\hline
$q$ & 2 & 3 & 4 \\
\hline
\hline
$t$ & $-7.1(3) \times 10^{-8}$ & $-9.1(2) \times 
10^{-9}$ & $-8(1) \times 10^{-10}$ \\
\hline
$\tau$ &  $-2.1(6) \times 10^{-8}$ & $-2.3(2) 
\times 10^{-9}$ & $- 2(77)\times 10^{-12}$  \\
\hline
$r$ &  $ 2.5(2) \times 10^{-9}$ & $2.8(1) 
\times 10^{-11}$ &  $1.3(5) \times 10^{-11}$ \\
\hline
\end{tabular}
\end{center}
\label{table:mom}
\end{table}
We see that for all the cases $\mu_{3}$ is smaller for scaling variable 
$r$,
with one exception $q=4$, $\tau$, when numerical error is approximately 
thirty
times greater than value $\mu_{3}(\tau)$ and six times greater than 
$\mu_{3}(r)$.
For all other lattice sizes $L=16,\dots,112$ the third moment  $\mu_{3}$ 
calculated by using variable $r$, is smaller than for $t$ or for $\tau$.
Therefore, we will work with the scaling variable $p=1-\exp(-\beta)$,
the probability of bond to be closed for $q=2,3,4$.
The critical point of the $q$-state Potts model 
on the infinite lattice is~\cite{Baxter} 
$p_{c}(q)=\frac{\sqrt{q}}{\sqrt{q}+1}$.
The bond percolation critical point $p_{c}(q=1)=\frac{1}{2}$
can be obtained from this formula. Variable
$p=1-\exp(-\beta)$  naturally provides crossover
from the percolation to the Potts model. We can plot the crossing
probability $\pi_{hs}$ as a function of $p$ for the percolation
and the Potts model on the same graph.

\begin{figure}[t]
\begin{center}
\epsfig{file=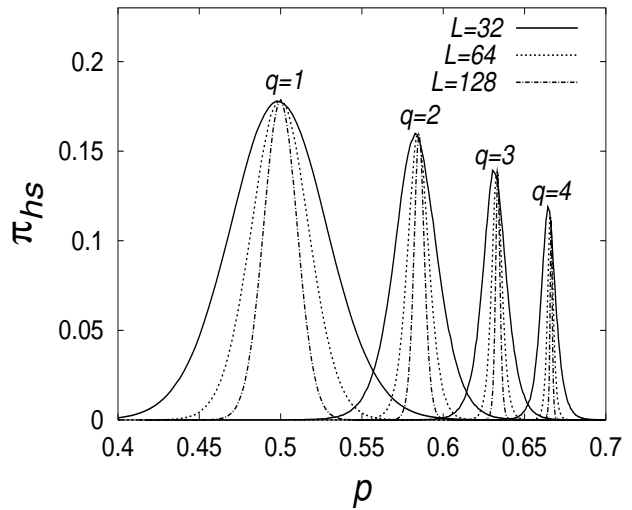,width=8.6cm,height=7cm}
\end{center}
\caption{The crossing probability $\pi_{hs}(p;L,q)$
 for $q=1,2,3,4$ and $L=32,64,128$}
\label{figphs}
\end{figure}

In Fig.\ref{figphs}, data for $\pi_{hs}(p;L,q)$
are plotted for $q=1,2,3,4$ and $L=32,64,128$.
We can see shift of the critical point $p_{c}(q,L)$ for finite lattice 
sizes $L$.
 We also see the change of the width of the function due to size 
scaling.
Let us try to identify the shape of $\pi_{hs}$ by analyzing the ratio
$\frac{\mu_{4}}{\mu_{2}^{2}}$. 
We know that at least for the  percolation,  function $\pi_{hs}$
 looks like the Gaussian function~\cite{vas} $ \sim A 
\exp(-x^{2})$, and for the Gaussian function we expect 
$\frac{\mu_{4}}{\mu_{2}^{2}}(Gaussian)=3$.
But, in  reality,  crossing probabilities are not Gaussian.
The moments ratio for function $\pi'_{h}(p)$ 
(the derivative of crossing probability $\pi_{h}(p)$ with respect to $p$), 
for the percolation model 
was found by Ziff $\frac{\mu_{4}}{\mu_{2}^{2}}(q=1)
\simeq 3.20(5)$~\cite{Ziff1} 
and more precisely  $\frac{\mu_{4}}{\mu_{2}^{2}}(q=1) 
\simeq 3.176(4)$~\cite{Ziff2} and by Hovy and Aharony
$\frac{\mu_{4}}{\mu_{2}^{2}}(q=1)
\simeq 3.174(25)$~\cite{HA}. 
The explanation of the non-Gaussian crossing probability shape
was first given by Berlyard and Wehr~\cite{BW} and was also discussed 
in Ref.~\cite{Wester}.
Newman and Ziff verify that the tails of the crossing probability
falls of as $\exp(-c (p-p_{c})^{4/3})$, therefore the tail of the 
distribution is 
characterized by the correlation length exponent~\cite{NZ1,NZ2}.
 However, as the  first approximation 
we introduce  an additional scaling index
$\zeta$ and  check  moment ratios for function  $g(x;\zeta)=A 
\exp(-x^{\zeta})$.
For this function $g(x;\zeta)$ the moment ratios are
\begin{equation}
\label{eq:mom}
\frac{\mu_{4}}{\mu_{2}^2}(\zeta)=
\frac{\Gamma(\frac{1}{\zeta})\Gamma(\frac{5}{\zeta})}
{\Gamma(\frac{3}{\zeta})^{2}}
\end{equation}
 We calculate moments  
for several values of $\zeta$ and put these data into the
first ($\zeta$) and second ($\frac{\mu_{4}}{\mu_{2}^2}(\zeta)$) 
 rows of the Table~\ref{table:momg}. The choice of values 
 $\zeta$
will be argued later.
Results of computation of the  moments ratio 
$\frac{\mu_{4}}{\mu_{2}^2}(q)$  for 
the crossing probability 
$\pi_{hs}(p;L,q)$ for   $L=32,128$ is 
placed in the 
fourth and fifth row of the Table~\ref{table:momg}.
We see that the moments ratio practically does not depend upon the lattice 
size $L$. We check this fact for others lattice sizes.
\begin{table}[h]
\caption{The ratio of moments $\frac{\mu_{4}}{\mu_{2}^{2}}$ for 
$g(x,\zeta)$ 
and $\pi_{hs}(p;L,q)$}
\begin{center}
\begin{tabular}{|c|c|c|c|c|}
\hline
\hline
$\zeta $ & 2 & 3/2 & 4/3 & 6/5 \\
\hline
\hline
$ \frac{\mu_{4}}{\mu_{2}^2}(\zeta)\;\;$ {\rm analytically } & 
3 & 3.76195 & 4.22219 & 4.7434 \\
\hline
\hline
$q$ & 1 & 2 & 3 & 4\\
\hline
$ \frac{\mu_{4}}{\mu_{2}^2}(q)\;\;${\rm numerically }$L=32$ &  3.145(6) & 
3.871(7) 
&4.577(14) & 5.28(2) \\
\hline 
$ \frac{\mu_{4}}{\mu_{2}^2}(q)\;\;${\rm numerically }$L=128$ & 3.18(2) & 
3.91(8) 
& 
4.56(2) & 5.30(4) \\
\hline
\end{tabular}
\end{center}
\label{table:momg}
\end{table}
 Moment ratios for $g(x;\zeta)$ slightly 
differ from moment ratios for $\pi_{hs}$  (this fact will be explained 
later), but we can try to approximate the crossing probability
by the function $g(x,\zeta)$ and then compare results of approximation 
with numerical data.

Below, we describe the fitting procedure.
As we can see from Fig.\ref{figphs}, there are  many nonuniversal scaling 
factors for $\pi_{hs}$: the amplitude of the crossing probability 
$A(L,q)$,  the finite size critical point $p_{c}(L,q)$,
which differs from $p_{c}(L=\infty,q)=\frac{\sqrt{q}}{\sqrt{q}+1}$,
the nonuniversal scaling factor $B(L,q)$, which provide 
the finite size scaling of the function, and 
the additional scaling index $\zeta(L,q)$.

\begin{figure}[h]
\begin{center}
\epsfig{file=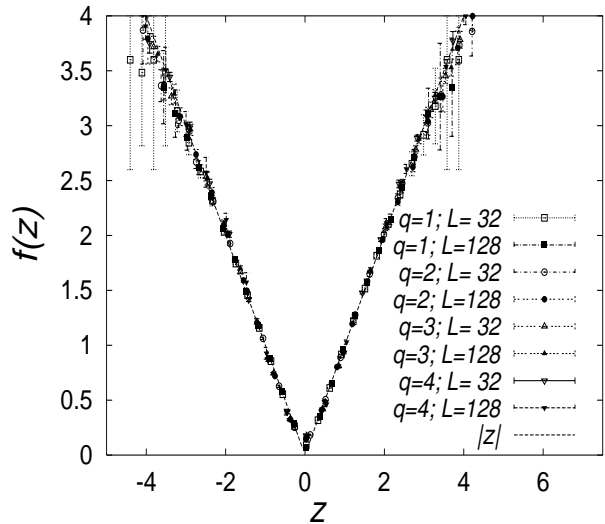,width=8.5cm,height=7cm}
\end{center}
\caption{The scaled
crossing probability $f(z)=|\log(Q(z))|=|\log(\pi_{hs})-\log(A(L,q))|$
as a function of the new scaling variable 
$z=[B(L,q)(p-p_{c}(L,q))]^{\zeta(L,q)}$
for $q=1,2,3,4$ and $L=32,128$.} 
\label{figscal1}
\end{figure}

 We perform the four-parametric fit of $\pi_{hs}(p;L,q)$
by the function~(\ref{eq:fitf})
\begin{equation}
\label{eq:fitf}
F(p;L,q)= A(q,L)\exp(  -  \left[ B(L,q)(p-p_{c}(L,q))  
\right]^{\zeta(L,q)}) 
\end{equation}
We use  points  $\log(\pi_{hs})>-9$
for this fit 
and the log scale for the 
ordinate axis. 
 As a result of the approximation we get 
a set of scaling amplitudes $A(q,L)$, nonuniversal metric  
$B(q,L)$, critical 
points $p_{c}(q,L)$ and scaling indices $\zeta(q,L)$.
Then,  we use this scaling factors to adjust our numerical 
data onto one line for 
$q=1,2,3,4$ and $L=32,128$.
We plot $f=|\log(\pi_{hs}(p;L,q))-\log(A(L,q))|$ 
as a function of the new scaling variable
$z= \left[B(L,q) \left(p-p_{c}(L,q) \right) \right]^{\zeta(L,q)}$
in Fig.\ref{figscal1}. 

The scaling straight lines $|z|$ shown in Fig.\ref{figscal1},
correspond to the  fitting function  $F(z)=\exp(-|z|)$.
We see that all points lie on the one curve,
and this curve is very close to $|z|$ in the range $0.1 < z <3$. So, we 
might expect that  our fitting procedure is correct.
 However, on the graph, the tail  points deviate from
the line $|z|$, and this deviation explains the fact that moment ratios
for $\pi_{hs}(p;L,q)$ in the Table~\ref{table:momg}  differ from 
analytical values  for $g(x,\zeta)$. 
Results of approximation for the crossing amplitude $A(L,q)$ are placed 
in Table~\ref{table:a}.
For each value of $q=1,2,3$ amplitudes $A(L,q)$ are equal
within our numerical accuracy of the approximation. Therefore, we can 
conclude that the scaling amplitude depends upon $q$
and  depends  weakly upon $L$.
\begin{table}[h]
\caption{The scaling amplitude  $A(L,q)$ for $L=16,32,64,128$}
\begin{center}
\begin{tabular}{|c|c|c|c|c|}
\hline
$q$ & 1 & 2 & 3 & 4\\
\hline
\hline 
$A(L=16)$ & 0.1806(2) &0.166(1) & 0.151(2)& 0.137(2) \\
\hline
 $A(L=32)$ & 0.1794(3) & 0.167(1) & 0.151(2)& 0.132(2) \\
\hline 
$A(L=64)$ & 0.1806(3)&  0.167(1) & 0.148(1) & 0.128(2) \\
\hline
$A(L=128)$ & 0.1794(3) & 0.166(1) & 0.149(2) & 0.122(2)\\
\hline
\end{tabular}
\end{center}
\label{table:a}
\end{table}

Results of the approximation for the scaling index $\zeta(L,q)$ are
represented in the Table~\ref{table:zeta}. We can see that
this scaling index practically does not depend upon lattice size.
\begin{table}
\caption{The scaling index $\zeta(L,q)$ for $L=16,32,64,128$}
\begin{center}
\begin{tabular}{|c|c|c|c|c|}
\hline
$q$ & 1 & 2 & 3 & 4\\
\hline
\hline 
$\zeta(L=16)$ & 1.921(3)  & 1.577(12) &1.398(15) & 1.264(18)\\
\hline
 $\zeta(L=32)$ & 1.900(4) & 1.557(12)  & 1.364(13) & 1.238(14)  \\
\hline 
$\zeta(L=64)$ &1.886(4) & 1.55(12)& 1.367(13) & 1.218(15)\\
\hline
$\zeta(L=128)$ & 1.887(5) &   1.545(14) & 1.337(13) &  1.198(14)\\
\hline
\end{tabular}
\end{center}
\label{table:zeta}
\end{table}
We assume that for each value $q$ we have the additional scaling index 
$\zeta(q)$, which is
nondependent upon the lattice size.  Test values of $\zeta$ in the
first row of Table~\ref{table:momg} are chosen to be close to 
$\zeta(q)$. It seems, that $\zeta(3)=4/3$, $\zeta(4)=6/5$.

\begin{figure}[h]
\begin{center}
\epsfig{file=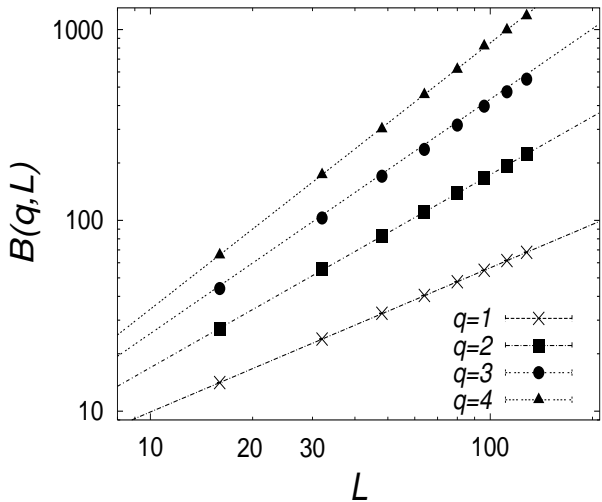,width=8.5cm,height=7cm}
\end{center}
\caption{Nonuniversal metric factors $B(L,q)$. Results of  approximation  
by the function $b(q)L^{y(q)}$ are shown by lines (see 
Table~\ref{table:b}).}
\label{figb}
\end{figure}  

We know that in accordance with the scaling theory
for each fixed value $q$ the crossing probability
must be a universal function of variable 
$x=L^{\frac{1}{\nu(q)}}(p-p_{c}(L,q))$.
Therefore, we approximate the numerical data for 
$B(L,q)$~(see~Fig.\ref{figb}) 
by  function $b(q)L^{y(q)}$ and represent results
in Table~\ref{table:b}. Results of
the  approximation $b(q)L^{y(q)}$ are also shown 
in  Fig.\ref{figb} by lines.
\begin{table}[h]
\caption{The approximation of the nonuniversal metric factors $B(L,q)$
by the function $b(q)L^{y(q)}$. The inverse correlation length index 
$\frac{1}{\nu(q)}$ is added for the comparison.}
\begin{center}
\begin{tabular}{|c|c|c|c|c|}
\hline
$q$ & 1 & 2 & 3 & 4\\
\hline
\hline
 $b(q)$ & 1.737(6) & 1.65(3)  & 1.51(3) & 1.38(3)  \\
\hline 
$y(q)$ & 0.756(8) &   1.011(3) & 1.218(16) &  1.39(6)\\
\hline
$\frac{1}{\nu(q)}$ & 0.75 & 1 & 1.2 & 1.5 \\
\hline
\end{tabular}
\end{center}
\label{table:b}
\end{table}
 We can see that the thermal scaling index $y(q)$ is very
close to the analytical value of the  inverse correlation 
length index $\frac{1}{\nu(q)}$~\cite{Baxter}, which is represented in 
the Table~\ref{table:b} for the comparison.
 This fact once more confirms our approximation procedure.
The exception is the case $q=4$, for which there is a difference
between $y(q=4)$ and $\frac{1}{\nu(q=4)}$. 

Many critical quantities in the Potts model $q=4$
exhibit logarithmic corrections~\cite{CNS,SS,AA,CZ}.
These logarithmic corrections explain the difference
between analytical value $\frac{1}{\nu(q=4)}=1.5$
and numerical approximation for the scaling index $y=1.39(6)$.

\section{Conclusions}

To summarize results, we can conclude
that the crossing probability $\pi_{hs}(p;L,q)$ is a 
universal function~$Q(z)$
\begin{equation}
\label{eq:uni}
\pi_{hs}(p;L,q)= A(q)Q \left(    \left[ 
b(q)L^{\frac{1}{\nu(q)}}(p-p_{c}(L,q))  
\right]^{\zeta(q)} \right) 
\end{equation}
of the scaling variable
$z=\left[ b(q)L^{\frac{1}{\nu(q)}}(p-p_{c}(L,q))
\right]^{\zeta(q)}$.
As we can conclude from  Fig.\ref{figscal1}, the function 
$Q(z)$ looks like the exponent $Q(z)\simeq \exp(-|z|)$
on the interval $0.1 < z <3$, but deviates from it in the vicinity of 0
and on the tails $|z|>3$.

Let us pay attention to the
 new details of this work.
We consider the universality of the crossing probability for 
different values of $q$ by adding the scaling index $\zeta (q)$. 
We work in the scale $p$, where
$p=1-\exp(-\beta)$ is the probability of a bond to be closed
instead of the usual scale $t=(T-T_{c})/T_{c}$ to make the crossing 
probability 
symmetric. We find numerically the scaling index  $\zeta(q)$. 
The universal function looks like $Q(z)\simeq \exp(-|z|)$
on the interval $0.1 < z <3$.

The author would like to thank Robert M. Ziff for helpful remarks, 
comments and discussion.
The author is  grateful to S. Nechaev for revision of the manuscript and 
to  the Joint SuperComputer Center RAS 
(www.jscc.ru) for
providing computational resources.

\end{document}